\documentclass{rsauthor}

\jname{Phil. Trans. R. Soc.}

\begin{document}

\title{High energy emission from transients}
\author{Hinton, J.A. \& Starling, R.L.C.}
\address{University of Leicester, University Road, Leicester LE1 7RH,
United Kingdom}
\date{}

\abstract{Cosmic explosions dissipate energy into their surroundings on a very wide range of time-scales: producing shock waves and associated particle acceleration. The historical culprits for the acceleration of the bulk of Galactic cosmic rays are supernova remnants: explosions on $\sim10^{4}$ year time-scales. Increasingly however, time-variable emission points to rapid and efficient particle acceleration in a range of different astrophysical systems. Gamma-ray bursts have the shortest time-scales, with inferred bulk Lorentz factors of $\sim$1000 and photons emitted beyond 100\,GeV, but active galaxies, pulsar wind nebulae and colliding stellar winds are all now associated with time-variable emission at $\sim$TeV energies. Cosmic photons and neutrinos at these energies offer a powerful probe of the underlying physical mechanisms of cosmic explosions, and a tool for exploring fundamental physics with these systems. Here we discuss the motivations for high-energy observations of transients, the current experimental situation, and the prospects for the next decade, with particular reference to the major next-generation high-energy observatory CTA.}

\keywords{transients; gamma-ray bursts; gamma-rays; neutrinos}


\maketitle

\section{Introduction}

Transient phenomena are often associated with the most extreme conditions in the Universe, where violent outbursts and explosions create jets and/or shocks in which particle acceleration can occur. Particles accelerated to high energies in these environments produce $\gamma$-rays that have been observed with high-energy instrumentation since the 1960's, but with a breakthrough in sensitivity associated with Energetic Gamma Ray Experiment Telescope (EGRET, see e.g. Thompson et al. 1995) observations in the early 1990's and the advent of sensitive ground-based instruments such as HESS (Hinton 2004) a decade later. 
The production mechanisms for $\gamma$-ray emission include the decay of neutral pions produced in strong interactions of accelerated protons and nuclei, Inverse Compton (IC) scattering of high-energy electrons on soft photon fields and Bremsstrahlung emission from electron encounters with ambient material. Two advantages of $\gamma$-ray emission with respect to more conventional probes of high-energy particles, such as synchrotron emission in the radio and X-ray bands, are that $\pi^{0}$-decay emission probes the (often) energetically-dominant accelerated protons and nuclei as well as electrons, and that in the case of electron accelerators IC emission allows unambiguous energy-density measurements, without assumptions such as equipartition with magnetic fields. The use of neutrinos, produced for example in charged pion decay, as a probe of high-energy transient phenomena is being pioneered using large-volume ice and water detectors such as IceCube (Abbasi et al. 2012).

Our current view of the GeV (10$^9$\,eV) sky as provided primarily by the Fermi satellite, is rich in time-variable phenomena with typical time-scales ranging from $\sim1$ second in $\gamma$-ray bursts (GRBs), to days in Galactic novae (e.g. V407 Cyg, Abdo et al. 2010), to years in the colliding-wind binary $\eta$-Carinae (Reitberger et al. 2012). Active galactic nuclei (AGN) are variable on a wide range of time-scales, with the shortest-time-scale variability probed only by the very-large collection area (ground-based) instruments available above $\sim50$\,GeV.  
TeV emission (10$^{12}$\,eV) is seen from cosmic explosions on time-scales up to thousands of years, the ages of the TeV-detected supernova remnants (SNRs), as well as minute to month time-scale variability from blazars, radio galaxies and Galactic binary systems (see e.g. Hinton \& Hofmann 2009). Many of these systems are `persistent', but the individual flaring events have much in common with transients such as GRBs.

Study of these time-variable phenomena at high energies, with the possibility to disentangle effects due to the cooling and transport of particles and the interplay between particles and magnetic fields, is a promising route to improve our 
understanding of the process of particle acceleration, and the role played by accelerated particles in cosmic explosions and the feedback these events have on their environments. 
There is great promise in this area in the future, with the construction of the Cherenkov Telescope Array (CTA) and a window on new types of transients with Adv. LIGO (Harry et al. 2010). In the nearer term the second phase of the HESS project will bring a major boost to $\sim$50\,GeV studies and the prospect for neutrino detection of transients with IceCube remains strong.

Here we discuss first the currently operating detectors, then the results from these instruments and finally the prospects for high-energy transient studies in the decades to come.

\section{Current Detectors} \label{sec:current}

The Fermi Gamma-ray Space Telescope, launched in 2008, is a two-instrument mission operating between 10\,keV and $\sim$200\,GeV. 
From a high-energy perspective Fermi's Large Area Telescope (LAT, Atwood et al. 2009), a pair-conversion telescope, is of most interest. LAT operates in the energy range above 25\,MeV, has a collection area of $\sim$1~m$^{2}$, and is currently the most sensitive instrument up to $\sim50$\,GeV, reaching a differential flux sensitivity (i.e. the flux required for detection in 4 independent spectral bins per decade in energy) of $\approx 10^{-12}$ erg cm$^{-2}$ s$^{-1}$ at $\sim$300\,MeV for a steady extragalactic point-like source in the current $\approx$4 year all-sky survey dataset (see Fig.~4 for the LAT sensitivity in comparison to the planned CTA observatory). The Gamma-ray Burst Monitor (GBM) acts as an all-sky monitor of hard X-ray transients.
Together GBM$+$LAT are revealing the high-energy spectra of numerous transient sources, with particularly important discoveries in the GRB field (Section~3), and in setting constraints on quantum gravity models by constraining Lorentz invariance violation (e.g. GRB\,090510, Abdo et al. 2009a). 

Two types of ground-based very-high-energy (VHE) instrument have been used to monitor the $\gamma$-ray sky. Air-shower particle detecting arrays such as MILAGRO (Atkins et al. 2003, and soon HAWC, e.g. Tepe et al. 2012 and see Section~4) have wide fields of view (FoV $\sim1$ sr) and very high duty cycle (close to 100\%) but modest resolution and collection area. Instruments based on the Imaging Atmospheric Cherenkov Technique (IACT) are more precise and achieve collection areas in excess of $10^{5}$ m$^{2}$ but have smaller FoV ($\sim$10 square degree) and operate only during darkness. The major current generation IACT arrays are MAGIC (e.g. Baixeras et al. 2004; Cortina et al. 2005), VERITAS (Holder et al. 2006) and HESS, all with multiple telescopes with large ($\ge12$ m diameter) segmented mirrors providing stereoscopic imaging of the electromagnetic cascades initiated by high-energy $\gamma$-rays in the Earth's atmosphere. 
IACT sensitivity, angular and energy resolution, and collection area can all be improved by adding more, or building larger, telescopes. 
Imaging atmospheric Cherenkov telescopes can and do slew to transient events, alerted by the Gamma-ray burst Co-ordinate Network, but are limited by duty-cycle and redshift depth is somewhat limited (especially at multi-TeV energies) due to the $\gamma$-ray opacity of the Universe (due to pair-production on the Extragalactic Background Light, EBL, see e.g. Stecker et al 1992). The performances of HESS, MAGIC and VERITAS as operating circa 2011 are very similar, with $\sim0.1^{\circ}$ resolution (an order of magnitude better than Fermi LAT at 1\,GeV) and differential flux sensitivity of a few $10^{-12}$ erg cm$^{-2}$ s$^{-1}$ at $\sim$1\,TeV for a 50 hour observation. Upgrades to these systems are however underway or have been recently completed. Of particular note is the second phase of the HESS project, which involves the addition of
a new 28~m diameter telescope (see Figure~1), bringing the capability to detect photons of much lower, $\sim 20$\,GeV, energies and improve sensitivity by a factor of two or more at 300\,GeV. July 2012 saw the first-light of HESS-2 and the first scientific observations are now underway. The new large telescope is designed to slew to any point in the sky in $\sim$40 seconds and transients form a key part of the science programme.
The science highlights from the current generation detectors include the census of Galactic particle acceleration sites, the detection of distant AGN with implications for the density of far IR photons in the universe, the discovery of high-energy emission from starburst galaxies and the detailed imaging of TeV emission from supernova remnants and pulsar wind nebulae (see Hinton \& Hofmann 2009 and references therein). The rapid variability of AGN in the TeV-blazar class is discussed in Section~3.

Beyond photons, very high-energy neutrinos and ultra-high-energy (UHE) protons are promising messengers with which to probe the extreme universe. The prime current facility for high-energy neutrino detection at present is the deep ice-Cherenkov detector IceCube. No neutrino sources have so far been identified and the limits derived are constraining models for high-energy phenomena including GRBs (see Section~3). The Pierre Auger Observatory (\u{S}m\'ida et al. 2012 and references therein) is the most sensitive detector of cosmic rays (and also neutrinos and photons) at energies above $\sim10^{19}$\,eV. 
So far no strong evidence for individual sources of UHE cosmic rays has been found and charged messengers such as protons are expected to experience long time delays due to longer path-lengths with respect to photons, making transient identification at UHE very difficult.

\begin{figure}
\includegraphics[width=0.5\linewidth]{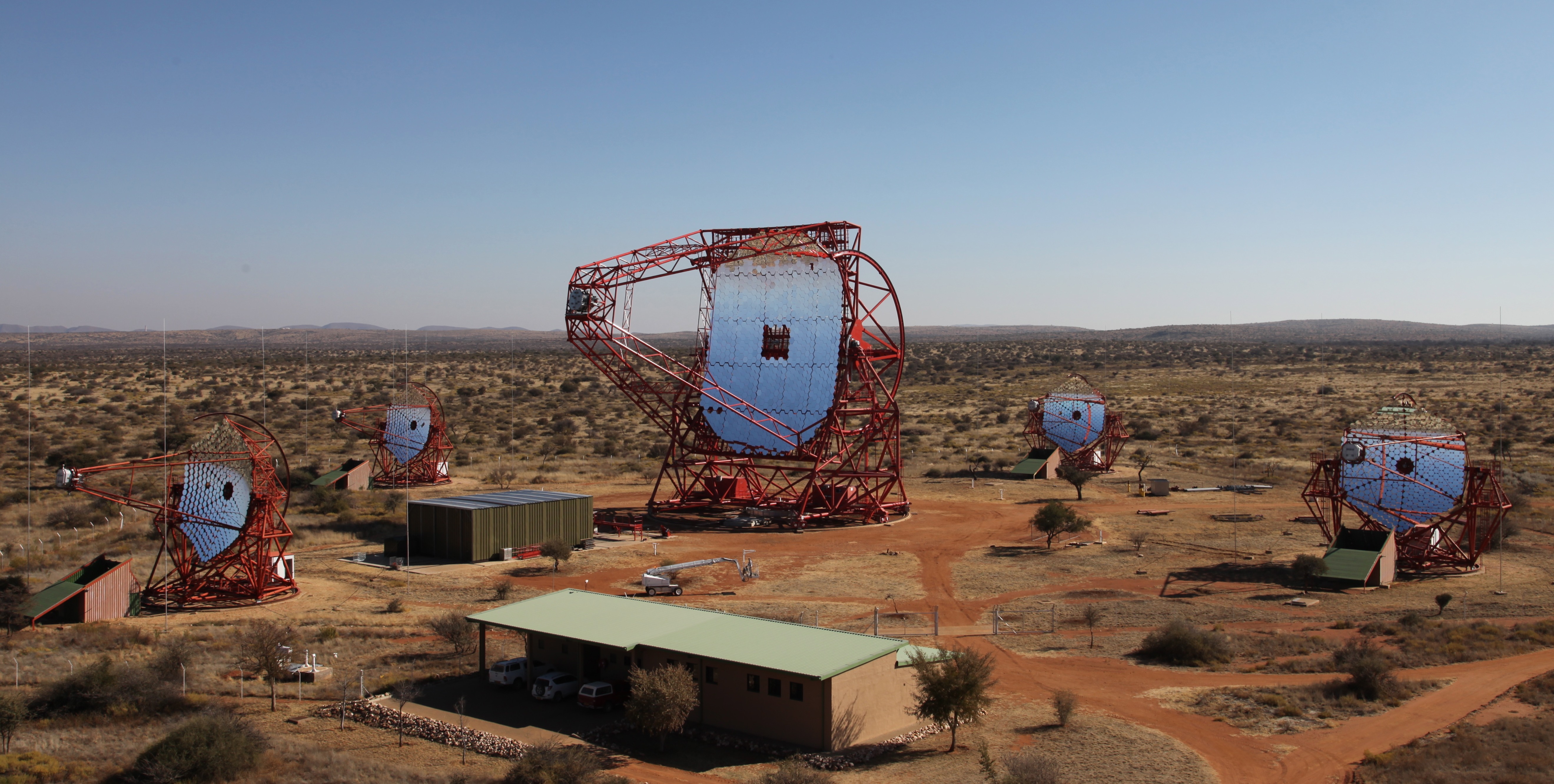} 
\includegraphics[width=0.5\linewidth]{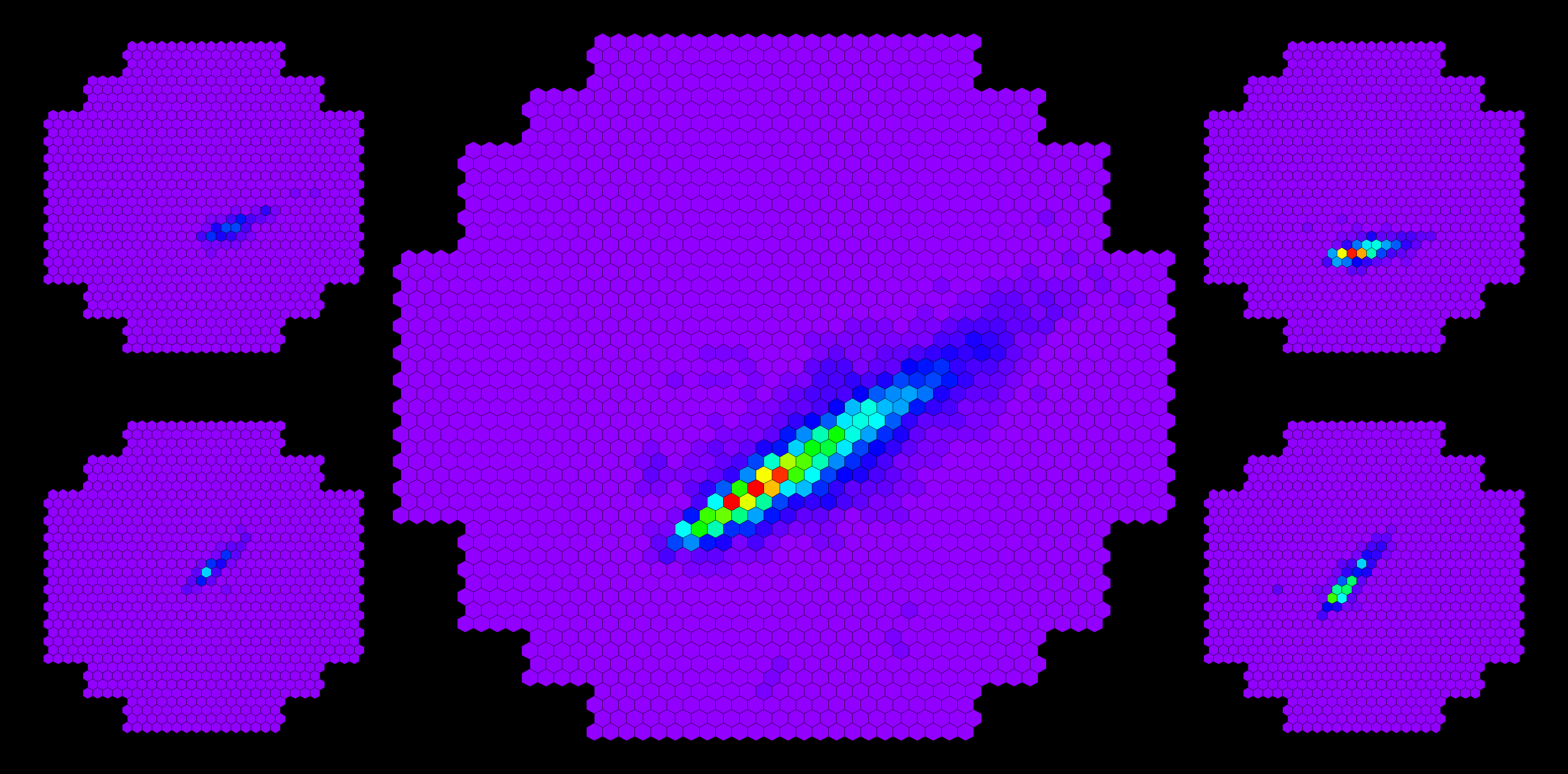}
\caption{Phase-2 of HESS in August 2012, together with one of the first air-shower Cherenkov events seen by all the cameras of the full 5-telescope system (reproduced with the permission of the HESS collaboration).}
\label{fig:hess}
\end{figure}

\section{High-Energy Results} \label{sec:GRBs}

\subsection{Gamma-ray Bursts}

GRBs are the most powerful transients in the Universe, releasing of order 10$^{51}$ erg of energy in shocks produced in a relativistic jet. The jets are formed during stellar collapse at the death of a massive star and the birth of a new black hole and supernova (collapsar); some GRBs may arise from compact object mergers. We are alerted to the onset of a GRB by a huge release of $\gamma$-rays, with the energy output typically peaking around 1\,MeV, lasting anywhere between one and several hundred seconds. Some of these events are now being recorded with Fermi to beyond 10\,GeV. 

Fermi LAT is detecting $\sim$10 GRBs yr$^{-1}$ at $>$100 MeV, and the highest-energy photons detected originated with energies of beyond 100\,GeV (GRB\,090902B at $z=1.822$, Abdo et al. 2009b). 

Both long and short GRBs, thought to be associated with collapsars and compact object mergers, respectively, have been observed at GeV energies. 
The mechanism by which this prompt emission is produced is not known. 
The initial GRB spectrum can be described by a Band function (Band et al. 1993) - a broken power law peaking (in $\nu F_{\nu}$) in the $\gamma$-ray band, possibly attributable to synchrotron emission. This may be augmented by a thermal or quasi-thermal high-energy `bump', and in some cases also a hard power law peaking beyond 100\,GeV and also of unknown origin (see Figure~2). 
GRB\,090926A was observed with LAT, and found to require an extra component at high energies to extend the emission beyond the reach of the Band function (Ackermann et al. 2011). This additional component also showed a cut-off at 1.4\,GeV which, if due to internal pair production, allows the Lorentz factor of the flow to be determined directly. This is a very exciting prospect for GRB physics. Two further LAT GRBs have high-energy components that show no clear cut-off (including short GRB\,090510, Ackermann et al. 2010), while an additional high-energy component is not required (but is poorly constrained) in other (fainter) examples. 
A delayed onset and longer duration of the highest energy emission compared to lower energy photons is also observed. Intriguingly, the delay appears to scale approximately with GRB duration, the long GRBs showing the greatest observed delays, irrespective of GRB luminosity (e.g. Toma, Wu \& M\'esz\'aros 2011). The long and short GRBs appear to differ only in their fractional radiative output at high energies - reduced in the long GRBs compared with the short GRBs, making VHE observations of short GRBs an exciting, but difficult, proposition.

Observations of GRBs at GeV to TeV energies can provide measurements of the bulk Lorentz factor and total energy budget, and help improve our understanding of the central engine, particle acceleration, jet formation, energy dissipation and GRB progenitors. Additionally, GRBs are unique probes of Lorentz Invariance Violation, and cosmology via EBL absorption signatures.
However, whilst Fermi has driven significant progress in high-energy observations of GRBs, ground-based instruments such as IACTs are yet to make VHE detections. This likely reflects the low rate of GRBs suitable for immediate observation by the ground-based facilities. The flux limits achieved at 1--10\,ks after burst triggers demonstrate that, despite absorption by the EBL, under more favourable conditions detections are clearly feasible (e.g. Aleksi\'c et al. 2010,2012). Indeed, in 2006 HESS was able to observe a Swift GRB-like trigger during the entire prompt phase (Aharonian et al. 2009), which occurred within the FoV of a pre-planned pointed observation. The trigger turned out to be a Galactic neutron star X-ray binary rather than a GRB (Wijnands et al. 2009).
Such comprehensive coverage on a GRB would be extremely valuable in constraining emission mechanisms for the prompt emission.

In the era of gravitational wave detection capability (with Advanced-LIGO/-VIRGO), GRBs will be at the forefront of efforts to locate an electromagnetic counterpart to a gravitational wave signal. It is important, therefore, to ensure facilities to detect, localise and follow-up GRBs and other potential counterparts are operational from $\sim$2015.

\begin{figure}
\includegraphics[width=1.0\linewidth]{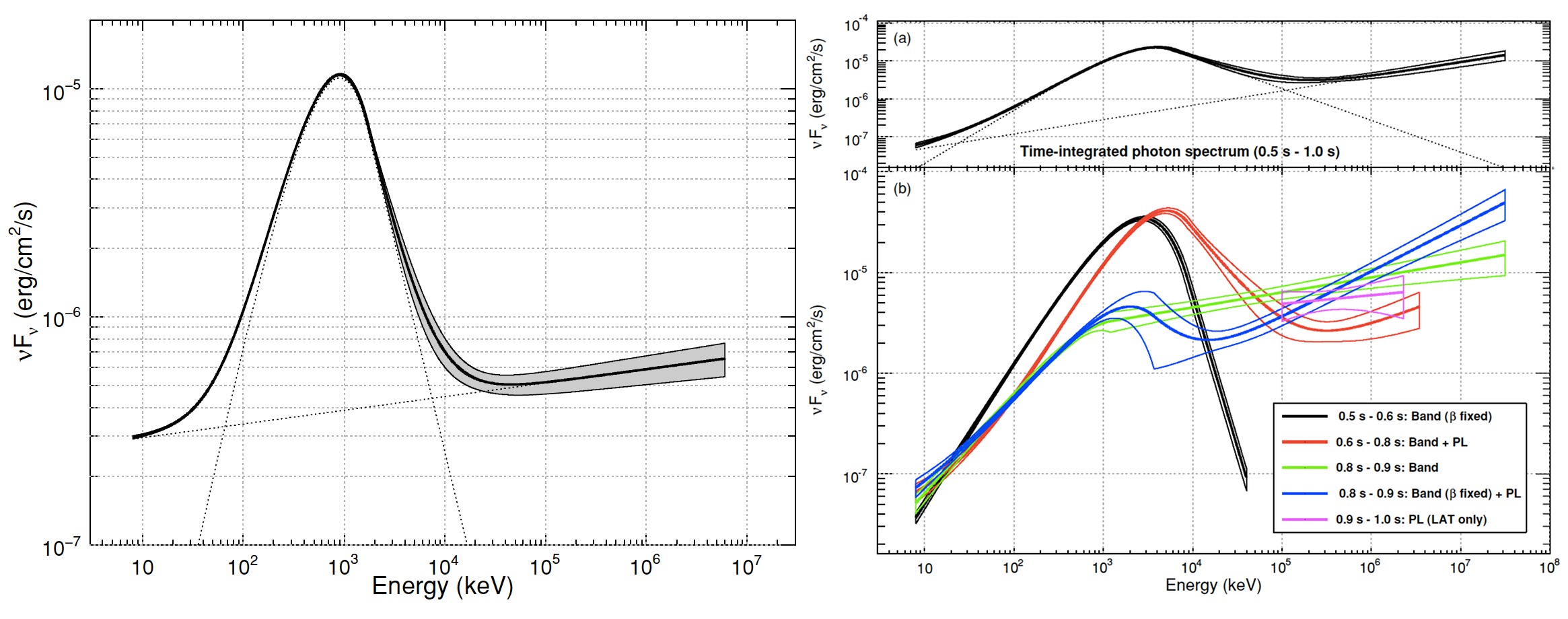}
\caption{Fermi LAT best-fit spectral models for long GRB\,090926A (left) and short GRB\,090510 (right), both requiring an additional, high-energy component (adapted from Ackermann et al. 2011; Abdo et al. 2009b). 
}
\label{fig:fermi}
\end{figure}

Results from GRB observations with IceCube are negative so far, with zero neutrinos detected from 300 GRBs (Abbasi et al. 2011). This result places the average neutrino flux a factor four below predictions from models where $p-\gamma$ interactions produce neutrons that escape and later decay to produce the observed flux of ultra-high-energy cosmic rays (UHECRs). GRBs are not, however, excluded as the sources of UHECR and more sensitive neutrino searches are needed to excluded scenarios with lower neutrino production efficiency.

\subsection{Other transient and rapidly variable phenomena}
Several other object classes with (relatively) short-time-scale variability have now been studied at high energies, including novae, binary systems with and without compact objects, AGN and pulsar wind nebula. The recent detection of dramatic variability in the high-energy emission from the Crab Nebula (Abdo et al. 2011; Tavani et al. 2011) and in particular the dramatic flare of April 2011 (Buehler et al. 2012),
suggests that acceleration of particles up to PeV energies is possible in this object on time-scales of $\sim$10 hours, with transient emission briefly dominating the flux from this object with a diameter of 10 light years. Whilst neither the point of origin within the Nebula nor the particle acceleration mechanism are currently understood, a matching new component above 10--100\,TeV  is expected (Kohri, Ohira and Ioka 2012) and can be probed by instruments such as CTA (Section~4). The processes at work in the Crab may extend to other ultrarelativistic shocks, such as those found in AGN jets, with important consequences for our understanding of these objects. Indeed very short-time-scale emission is seen in the TeV band from active galaxies with jets closely aligned with the line-of-sight of the observer: the so-called {\it blazars}. The flare from the prominent X-ray and TeV $\gamma$-ray blazar PKS\,2155$-$304 shown in Figure~3 is the most dramatic example to date, a factor $>$100 flux increase with respect to the quiescent state and variability on time-scales close to 1~minute. Such events indicate extremely large bulk Lorentz factors for the emission sites within the AGN jet, approaching the extreme values seen for gamma-ray bursts (Begelman, Fabian \& Rees 2008). Variability is also seen in objects where jets are not closely aligned with the line-of-sight to the observer, with the day-time-scale TeV variability of M\,87 (Aharonian et al. 2006) as the best studied case. Simultaneous VLBI and TeV observations of M\,87 indicate a strong increase in flux from the nucleus during VHE flares (Acciari et al. 2009), suggesting particle acceleration is taking place very close to the central supermassive black hole.

\begin{figure}
\vspace{-10cm}
\includegraphics[width=1.0\linewidth]{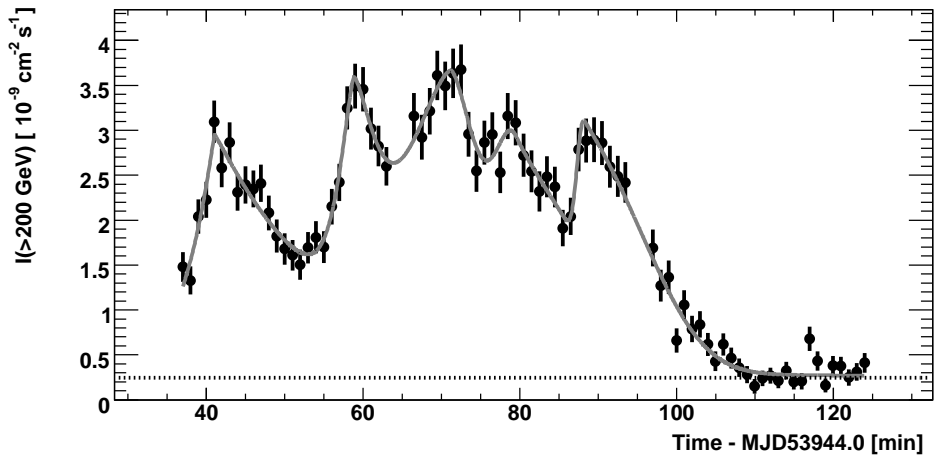}
\caption{
Minute-time-scale variability in the very-high-energy emission of the blazar PKS\,2155$-$30. The dotted horizontal line indicates the flux from the Crab Nebula, the brightest steady source in the sky at these energies (reproduced from Aharonian et al. 2007).
}
\label{fig:2155}
\end{figure}

\section{Future Prospects}\label{sec:future}

In general the prospects for the very high-energy and multi-messenger astronomy of transient phenomena seem very bright. So far IceCube has produced only upper limits for GRBs but the exposure is steadily increasing and future nearby GRBs would produce very constraining limits if no neutrinos are observed. The larger volume of the planned KM3NeT detector (Biagi et al. 2012) may also help to reach the critical threshold for transient astronomy with high-energy neutrinos.

Gamma-ray observations are perhaps the most exciting area in the near future. Beyond the currently operating detectors described in Section~2, several important new facilities are under construction or being planned.
HAWC (e.g. Tepe et al. 2012) will provide continuous coverage of a steradian of sky at TeV energies, providing much greater sensitivity to transient phenomena (see Abeysekara et al. 2012) in comparison to its predecessor MILAGRO due to its higher altitude site and larger detector area. Completion of HAWC is expected by $\sim$2014. The major current ground-based initiative in this area is however CTA -- the Cherenkov Telescope Array. CTA will bring a huge improvement in all aspects of performance with respect to current IAC systems, including the best energy flux sensitivity and angular resolution ($\sim$1$'$ at 10\,TeV) above the X-ray domain. The expected field of view is 8$^{\circ}$. It will comprise $>$100 telescopes of 3 different sizes, some of which will be fast-slewing (to anywhere on the sky in less than a minute), located at two sites. A 185M Euro European-led project, it builds on expertise gained from HESS, MAGIC and VERITAS among 170 institutes in 27 countries. As Figure~4 shows, the sensitivity of CTA is orders of magnitude better than Fermi for minute-time-scale phenomena.

\begin{figure}
\begin{centering}
\vspace{-2mm}
\includegraphics[width=0.8\linewidth]{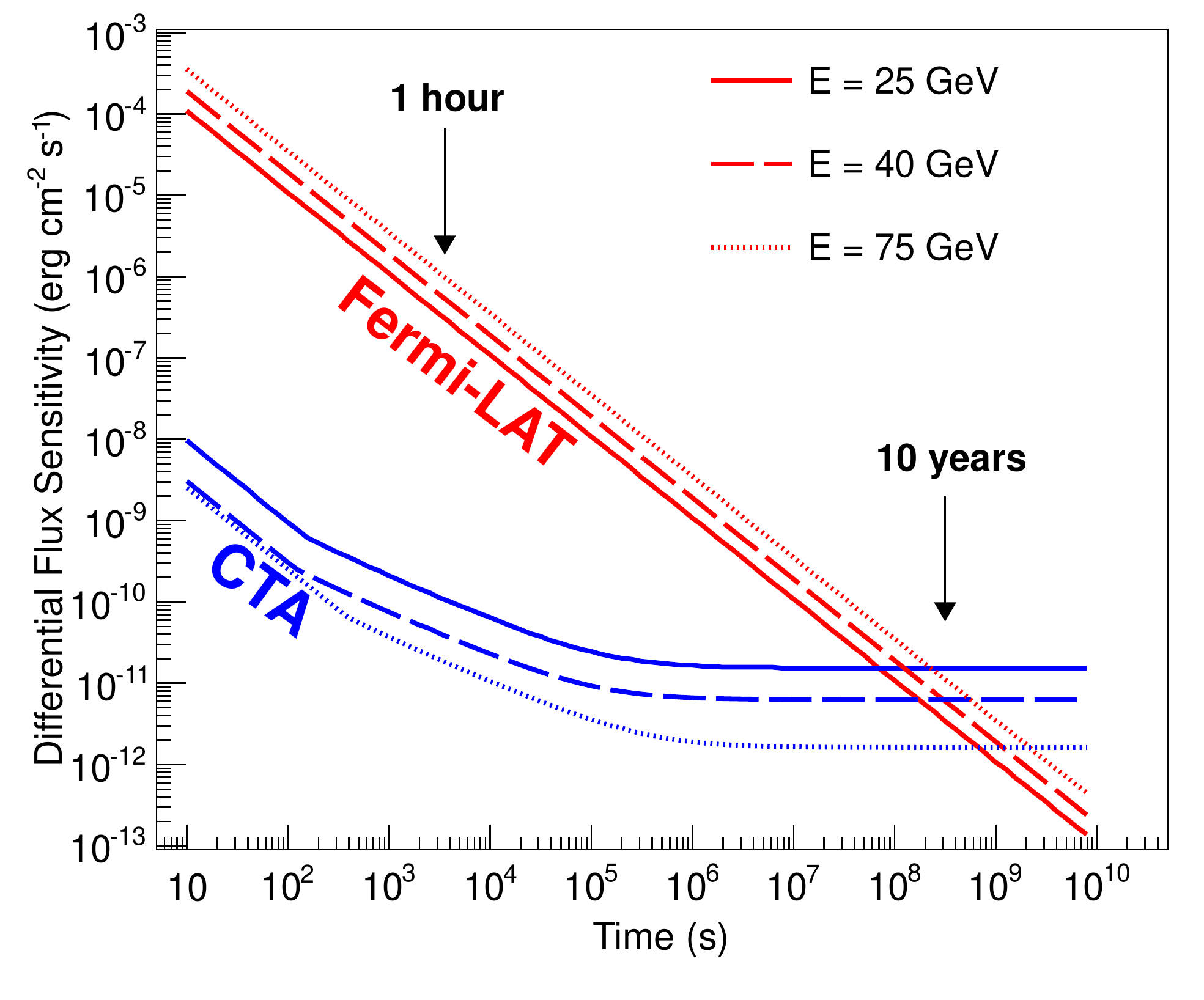}
\vspace{-5mm}
\caption{A comparison of the sensitivity of CTA and Fermi LAT as a function of integration time-scale, adapted from Funk \& Hinton 2013.}
\end{centering}
\label{fig:sens}
\end{figure}

CTA will both find and follow transients with its slewing capabilities and in a survey mode (see the Astroparticle Physics CTA Special Issue, including Inoue et al. 2013).
The transient follow-up program will rely on alerts from contemporaneous X-ray/$\gamma$-ray observatories, currently including Fermi and Swift (Gehrels et al. 2004), as well as from TeV (for example HAWC) and gravitational wave facilities. In addition, rapid photometry and redshifts from optical/near-infrared telescopes will be required to maximise the scientific impact of these observations. 

CTA will also generate its own triggers and alert multi-wavelength follow-up facilities to new transients discovered. A wide-field survey mode provides the best chance of serendipitous
transient detection (for e.g. short GRBs), which could equal the instantaneous HESS sensitivity
but over a much larger $\sim$30$^{\circ}\times30^{\circ}$ patch of sky. CTA surveys (Dubus et al. 2013) will have direct synergies with radio, optical and X-ray surveys, particularly with new facilities such as LOFAR, SKA and Advanced LIGO.

CTA will conduct a census of particle acceleration
across the universe. It can survey $\sim$400 times faster than HESS, and the improved resolution avoids source confusion and aids multi-wavelength follow-up.
With CTA we will uniquely probe stellar birth and death and cosmic ray
feedback both in our own galaxy and up to galaxy-cluster scales.

\section{Summary}
 
The transient and variable universe is taking on increasing importance in astrophysics, with rapidly improving facilities for very wide field monitoring across, and beyond, the electromagnetic spectrum. Observations at very high energies are no exception, with major new facilities planned and enormous potential for transient related science. The acceleration of high-energy particles seems to be a common feature of cosmic explosions and new observations at GeV energies and above, in photons and in neutrinos, offer powerful new probes of this phenomenon and of fundamental physics. CTA stands out as the major instrument in this area in the near future, and a suite of multi-wavelength instruments is required to fully exploit it. In particular future wide-field X-ray/soft-gamma-ray missions are vitally important for the future of high-energy transient astrophysics.

\ack{JAH acknowledges support from the Leverhulme Trust and RLCS acknowledges support from a Royal Society Fellowship. The authors are grateful to Paul O'Brien for presenting on their behalf.}

\end{document}